\documentstyle[epsfig,referee]{mn}
\begin{document}
    \halign{#\hfil         \cr
           } 
\title{Statistical Interpretation of LMC Microlensing Candidates.
}
\author [Sohrab Rahvar ]
{ Sohrab Rahvar $^{1,2}$ \\
$^1$ Department of Physics, Sharif University of Technology,\\
 P.O.Box 11365--9161, Tehran, Iran\\
$^2$ Institute for Studies in Theoretical Physics and Mathematics,
P.O.Box 19395--5531, Tehran, Iran }

 \maketitle


\begin{abstract}
After a decade of gravitational microlensing experiments, a dozen
of microlensing candidates in the direction of the stars of the
Large Magellanic Cloud (LMC) have been detected by the EROS and MACHO 
groups. Recently it was shown that the
distribution of the duration of the observed LMC microlensing
events is significantly narrower than what is expected from the
standard halo model. In this article we make
the same comparison, using non-standard halo models and considering 
the contribution of non-halo components of the Milky Way such as the disc, 
spheroid and LMC itself in the microlensing events. Comparing the 
theoretical and experimental widths of distributions of the duration of 
events shows that neither standard nor non-standard halo models are 
compatible with the microlensing data at least with $95$ per cent of 
confidence. This results maybe explained if (i) the MACHOs in the Galactic 
halo reside in clumpy structures or (ii) the durations of events
have been underestimated due to the blending effect.
\keywords: Galaxy: halo -- dark matter 
\end{abstract}
\section{Introduction}
The rotational curve of disc in the spiral galaxies shows the
existence of dark matter in the halos of galaxies. MAssive Compact
Halo Objects (MACHOs) like brown dwarfs, white dwarfs, neutron
stars and black holes could be candidates for the baryonic
component of the dark halo. The gravitational microlensing
technique was proposed by Paczy\'nski (1986) for indirect
detection of MACHOs in the halo of our Galaxy. The effect of
microlensing is a temporary light amplification of background stars
due to MACHOs passing through our line of sight. Since early 1990s
several groups like AGAPE, DUO, EROS, MACHO, OGLE and PLANET have
contributed to this field and began monitoring millions of stars
in the Large and Small Magellanic Clouds (LMC and SMC). In the
direction of LMC, the MACHO\footnote{http://wwwmacho.mcmaster.ca/}
collaboration observed $13-17$ candidates from $5.7$ years
observation of $11.9$ million stars (Alcock et al. 2000).
EROS\footnote{http://eros.in2p3.fr/} also observed LMC-1 from EROS
I photometric plates (Ansari et al. 1996) and the other four events
from EROS II (Lasserre  et al. 2000; Spiro \& Lasserre 2001;
Milsztajn \& Lasserre 2001). Interpretation of the observed results
within the framework of Galactic models has been a matter of debate
in recent years. By comparing the expected and the observed numbers of
microlensing events, it is possible to evaluate the mass
fraction of the halo in the form of MACHOs and also the mean mass of MACHOs.\\
As an example, in the standard Galactic model, MACHO group
obtained the optical depth of microlensing events $\tau_{LMC} =
 1.2^{+0.4}_{-0.3} \times 10^{-7}$ in the direction of LMC
(Alcock et al. 2000) and this result is consistent with the 
theoretical expectation if $ \sim 0.2 $ times the halo mass in 
this model is made up of MACHOs. The mean value of the duration of 
events also indicates that
the mean mass of lenses should be $\sim 0.5 M_{\odot}$, which
means that the masses of MACHOs are about the same as those of 
white dwarf stars. The EROS group also put a constraint on the 
fraction of halo in
the form of MACHOs, with the masses of lenses in the range of
$10^{-7}M_{\odot}$ to $4 M_{\odot}$, excluding that no more than
$40$ per cent of the standard halo is made of objects with up to one
solar mass at $95$ per cent confidence (Spiro \& Lasserre 2001). \\
It should be mentioned that the conclusions of EROS and MACHO groups on 
the contribution of MACHOs to the mass of the dark halo and the
mean mass of lenses is in some cases at variance with other
observations. The outline of this contradiction is as follows 
(Gates \& Gyuk 2001).
\begin{itemize}
\item To allow the mass of the MACHOs to be in the range proposed
by microlensing observations, the initial mass function of MACHO 
progenitors of the Galactic halo should be different from the disc (Adams \&
Laughlin 1996; Chabrier, Segretain \& Mera 1996). Limits on the
initial mass function of the halo arise from both 
low- and high-mass stars. Low 
mass stars $(M<1M_{\odot})$ should still be active and visible, and heavy
stars $(M>8 M_{\odot})$ have evolved into Type II supernovae and 
have ejected heavy elements into the interstellar medium.
\item If there were as many white dwarfs in the halo as suggested by
microlensing experiments they would increase the abundance of
heavy metals via Type I Supernova explosions. Canal, Isern $\&$
Ruiz-Lapuente (1997) used this phenomenon and showed that halo
fraction in the white dwarfs has to be less than $5--10$ per cent. To
be compatible with the gravitational microlensing results, 
they proposed that the star-formation process in the halo is
possibly different from the local observations for single as well
as binary stars.
\item Recently Green \& Jedamzik (2002) mentioned that the observed 
distribution of microlensing duration is not compatible with what is 
expected from the standard halo 
model. They showed that the distribution of microlensing candidates in terms 
of the duration of events is significantly narrower compared to that  
expected from the standard halo model at $90-95$ per cent confidence.
\end{itemize}
Here we extend the earlier work of Green \& Jedamzik (2002) (i) to include the 
EROS 
microlensing events (ii) to take into account the contribution of the LMC 
and luminous 
components of the Milky Way in the microlensing events, and (iii) and 
finally to consider the non-standard halo models (Alcock et al. 1996). 
The mass function of lenses was chosen to be a Dirac-delta function where 
the peak of the function and the fraction of halo in the form of MACHOs in 
each Galactic model are chosen according to the likelihood analysis 
of the MACHO group. We generate the expected distribution of events, using 
the observational efficiency of the experiments and compare them with the
distributions of microlensing data. This comparison is performed
by a Monte-Carlo simulation to generate the width and the mean of 
distribution and compare them with the observed data.
The paper is organized as follows. In Section 2 we give a brief
review of Galactic models and generate the expected distribution of 
events by considering the EROS and MACHO observational efficiency. 
Section 3 compares the expected distribution in the Galactic models 
and observed data using statistical parameters. The results are 
discussed in Section 4.
\section{galactic models and the expected microlensing distribution}
This section discusses the relevant components of the structure of the
Milky Way, including: the power-law models of the Milky Way halo, 
luminous parts of Milky Way such as the disc and spheroid, and also the 
LMC itself. These elements can be combined to build various 
Galactic models that have been discussed by Alcock et al. (1996). 
Here we use the mass function of the MACHOs and their contribution 
to the mass of Galactic halo according to the likelihood 
analysis of the MACHO group. We obtain the theoretical distribution of 
the rate of events in the direction of the LMC in each model. The 
observational efficiencies of the EROS and MACHO experiments are applied to 
obtain the expected distribution of events as a function of the duration 
of events in these models.
\subsection{Power-law halo models}
Here we use the largest known set of axisymmetric models of
Galactic halo, the so-called "power-law Galactic" models (Evans 1994).
The density of these models in the cylindrical coordinate system
are given by
\begin{equation}
\rho(R,z)=  \frac{{V_a}^2{R_c}^{\beta}}{4\pi G q^2}\times
\frac{{R_c}^2(1+2q^2) + R^2(1-\beta q^2) +
z^2[2-(1+\beta)/q^2]}{({R_c}^2 + R^2 + z^2/q^2)^{(\beta+4)/2}},
\label{rho}
\end{equation}
where $R^2 = r^2 + z^2$, $R_c$ is the core radius and $q$ is the
flattening parameter which is the axial ratio of the concentric
equipotentials. $q=1$ represents a spherical $(E0)$ halo and $q
\sim 0.7$ gives an ellipticity of about $E6$. The parameter
$\beta$ determines whether the rotational curve asymptotically
rises, falls or is flat. At asymptotically large distances from
the centre of the Galaxy in the equatorial plane, the rotation
velocity is given by $V_{circ}\sim R^{-\beta}$. Therefore $\beta =
0$ corresponds to a flat rotation curve, $\beta<0$ is a rising
rotation curve and $\beta>0$ is a falling curve. The parameter $V_a$
determines the overall depth of the potential well and hence gives
the typical velocities of objects in the halo. The velocity
dispersion of halo also is given by
\begin{eqnarray}
\sigma_R^2 &=& \sigma_z^2 = \frac{V_a^4R_c^{2\beta}}{8\pi G 
q^2}\frac{2q^2R_c^2 +
(1-\beta)q^2R^2 + [2-(1+\beta)q^{-2}]z^2}{(1+\beta)(R_c^2 + R^2+z^2q^{-2})^{\beta+2}}\\
\sigma_{\phi}^2 &=& \frac{V_a^4R_c^{2\beta}}{8\pi G
q^2}\frac{2q^2R_c^2 + [2+2\beta- (1+3\beta)q^2]R^2 +
[2-(1+\beta)q^{-2}]z^2}{(1+\beta)(R_c^2 +
R^2+z^2q^{-2})^{\beta+2}}.
\end{eqnarray}
The parameters of power-law halo models are indicated in Table 1.
\subsection{Luminous components of the Milky Way and LMC}
The luminous and non-halo components of the Milky Way are the galactic
disc and spheroid. Here we also use the contribution of the LMC 
disc and halo. We model the density of the thin and thick discs of
the Milky Way and LMC as double exponentials (Binney \& Tremaine 1987):
\begin{equation}
\rho(R,z) = \frac{\Sigma_{0}}{2h} \exp\left[ -\frac{R -
R_{0}}{R_d} \right] \exp\left[-\frac{|z|}{h}\right]
\end{equation}
where $z$ and $R$ are cylindrical coordinates, $R_{0}$ is the
distance of the Sun from the centre of the Galaxy, $R_d$ is 
the scalelength, $h$ is the scaleheight of the disc and 
$\Sigma_{0}$ indicates the column density of the disc. 
For the thin disk of the Milky Way, which mainly consists of the 
star population 
and gases, these parameters are: $R_d = 4 kpc$, $h = 0.3 kpc$, 
$\Sigma_0 = 50 M_{\odot}pc^{-3}$, $R_0=8.5 kpc$  and $\sigma_v = 31 
km/s$, where $\sigma_v$ is the adopted one-dimensional velocity 
dispersion perpendicular to our line of sight. For the case of 
maximal disk, all the parameters are the same as the thin disk except 
$\Sigma_0 = 80 M_{\odot}pc^{-3}$. For the thick disc of the Milky Way, the 
parameters are: $R_d = 4 kpc$, $h =1.0 Kpc$, $\Sigma_0 = 40
M_{\odot}pc^{-3}$, $R_0 = 8.5 kpc$ and $\sigma_v = 49 km/s$. The
mass function of the disc component is taken according to the 
observations with the {\it Hubble Space Telescope} (Gould, Bahcall \& 
Flynn 1997). Here we are also
interested in considering the rate of microlensing by the LMC, the
so-called self-lensing. The LMC disc parameters taken
from Gyuk {\it et al.} (2000) are $R_d = 1.57 kpc$, $h = 0.3 kpc$, 
$\sigma_v = 25 km/s$.\\
The other luminous component that may have a contribution to the
microlensing events is the Milky Way spheroid. The spheroid density 
is given by (Guidice {\it el al.} 1994; Alcock {\it et al.}
1996):
\begin{equation}
\rho_{spher} = 1.18\times 10^{-4}(r/R_0)^{-3.5}M_{\odot}pc^{-3},
\end{equation}
This density profile clearly must be cut off at small distances
from the center of Galaxy, but this is irrelevant here since the
LMC line of sight is always at $r>0.99R_0$. We take the dispersion
velocity for this structure $\sigma_v = 120 km/s$.
\subsection{Expected rate of events in the Galactic models}
In this part we use the Galactic models to obtain the rate of
the duration of events.
To obtain the differential rate of duration of events we need entire
phase space distribution function. The differential rate is give by
\begin{equation}
d\Gamma = \frac{1}{m}F({\bf v},{\bf x})cos\theta u_T R_E v_t d^3v dx d\alpha,
\end{equation}
where $m$ is the mass of the lenses, $F(\bf{v},\bf{x})$ is the phase space 
distribution of the MACHOs, $u_T R_E$ is the radius of the microlensing 
"tube" at position $x$ from the observer, $u_T R_E d\alpha$ is the 
cylindrical segment of that tube and $v_t$ is the transverse velocity of 
the MACHO in the frame of the microlensing tube (Griest 1991). We use 
numerical methods to obtain the differential rate of events in the
standard halo model, power-law halo models and also in the disc, spheroid 
and LMC (Alcock et al. 1995). The Contributions of the
components of the Milky Way and LMC to the total differential rate 
of events are given by:
 \begin{equation}
\frac{d\Gamma}{dt} = {\it f}\frac{d\Gamma}{dt}(MWhalo) +
\frac{d\Gamma}{dt}(disk) + \frac{d\Gamma}{dt}(Spheroid)+
\frac{d\Gamma}{dt}(LMC),
\end{equation}
The first term is the halo contribution in which 
{\it f} is the halo fraction in the form of MACHOs. Second, third and 
fourth terms are the contributions of the disc, spheroid and LMC itself.
\begin{table*}
\begin{center}
\begin{tabular}{|c|c|c|c|c|c|c|c|c|c|c|}
\hline
\hline
& & Model :                & $S$    & $A$   & $B$  & $C$   & $D$ & $E$ & $F$ & $G$\\
\hline\\
&(1)& Description & Medium & Medium& Large& Small & E6 &
Maximal disk& thick disk & thick disk \\
&(2)& $\beta$              & --    &      0 & -0.2 &   0.2 &    0&    0&    0& 0\\
&(3)& $q$                  & --    &     1  & 1    &    1  & 0.71
& 1& 1& 1\\
&(4)& $v_a (km/s)$         & --    & 200    & 200  & 180 & 200
& 90 & 150 & 180 \\
&(5) & $R_c (kpc)$         & 5    & 5  & 5   &  5  &  5   & 20 &
25 & 20\\
&(6)& $R_0 (kpc)$          & 8.5  & 8.5& 8.5 & 8.5 & 8.5& 7 & 7.9
& 7.9 \\
&(7)& $\Sigma_0 (M_{\odot}/pc^{2})$ & 50 & 50 & 50 & 50 & 50 & 80
& 40 & 40 \\
&(8)& $R_d (kpc)$          &    3.5 & 3.5 & 3.5 & 3.5 & 3.5 & 3.5& 3 & 3 \\
&(9)& $z_d (kpc)$          &    0.3 & 0.3 & 0.3 & 0.3 & 0.3 & 0.3
& 1 & 1 \\
\hline
\end{tabular}
\label{tmodel}
\end{center}
\caption{The parameters of eight Galactic models. The first
line indicates the description of the models; the second line, the slope 
of the rotation curve $(\beta = 0$ flat, $\beta<0$ rising and $\beta>0$
falling); the third line, the halo flattening ($q =1$ represent
spherical); the fourth line, $(v_a)$, the normalization velocity; the 
fifth line, $R_c$, the halo core; the sixth line, the distance of the Sun 
from the centre of the Galaxy; the seventh line, the local column density 
of the disc ($\Sigma_0 = 50$ for canonical disc, $\Sigma_0 = 80$ for 
a maximal thin disc and $\Sigma_0 = 40$ for a thick disc); the eighth 
line, the disc scalelength; and the ninth line, the disk scalehight.}
\end{table*}
The parameter {\it f} and the mass function of MACHOs, which are taken to 
be a $\delta$-functions, can be obtained by the likelihood analysis 
method. Here we use the results of Alcock et al. (2000) for the $S,
B $ and $F$ models and that of Alcock et al. (1997) for $A,
C, D, E$ and $G$ models as shown in Table 2.
\begin{table*}
\begin{center}
\begin{tabular}{|c|c|c|c|c|c|}
\hline \hline
&Events   & Model & Halo    & $m_{ML} (M_{\odot})$ & $f_{ML}$  \\
\hline
&13       & S     & medium  & 0.54              & 0.20   \\
&17       & S     & medium  & 0.72              & 0.22   \\
&6        & A     & medium  & 0.32              & 0.41   \\
&8        & A     & medium  & 0.32              & 0.55   \\
&13       & B     & large   & 0.66              & 0.12   \\
&17       & B     & large   & 0.87              & 0.14   \\
&6        & C     & small   & 0.21              & 0.61   \\
&8        & C     & small   & 0.21              & 0.83   \\
&6        & D     &  E6     & 0.31              & 0.37  \\
&8        & D     &  E6     & 0.31              & 0.50  \\
&6        & E     &max disk & 0.04              & 2.8  \\
&8        & E     &max disk & 0.04              & $>1$   \\
&13       & F     & big disk& 0.19              & 0.39  \\
&17       & F     & big disk& 0.25              & 0.44  \\
&6        & G     & big disk& 0.21              & 0.71  \\
&8        & G     & big disk& 0.20              & 0.97  \\
 \hline
\end{tabular}
\label{tmodel}
\end{center}
\caption{ The maximum likelihood estimates of MACHO mass m, halo fraction
${\it f}_{ML}$, for the eight Galactic models are shown in this table. The 
first column shows the number of detected
microlensing events; the second column indicates the eight Galactic 
models, described in Table 1; the specifications of the models are
given in the third column; and the fourth and fifth columns show the
results of maximum likelihood analysis for the mass of MACHO and
halo fraction in the form of MACHOs.}
\end{table*}
The results of numerical calculations for the rate of events are 
shown in Fig. 1.
\begin{figure}
\begin{center}
\psfig{file=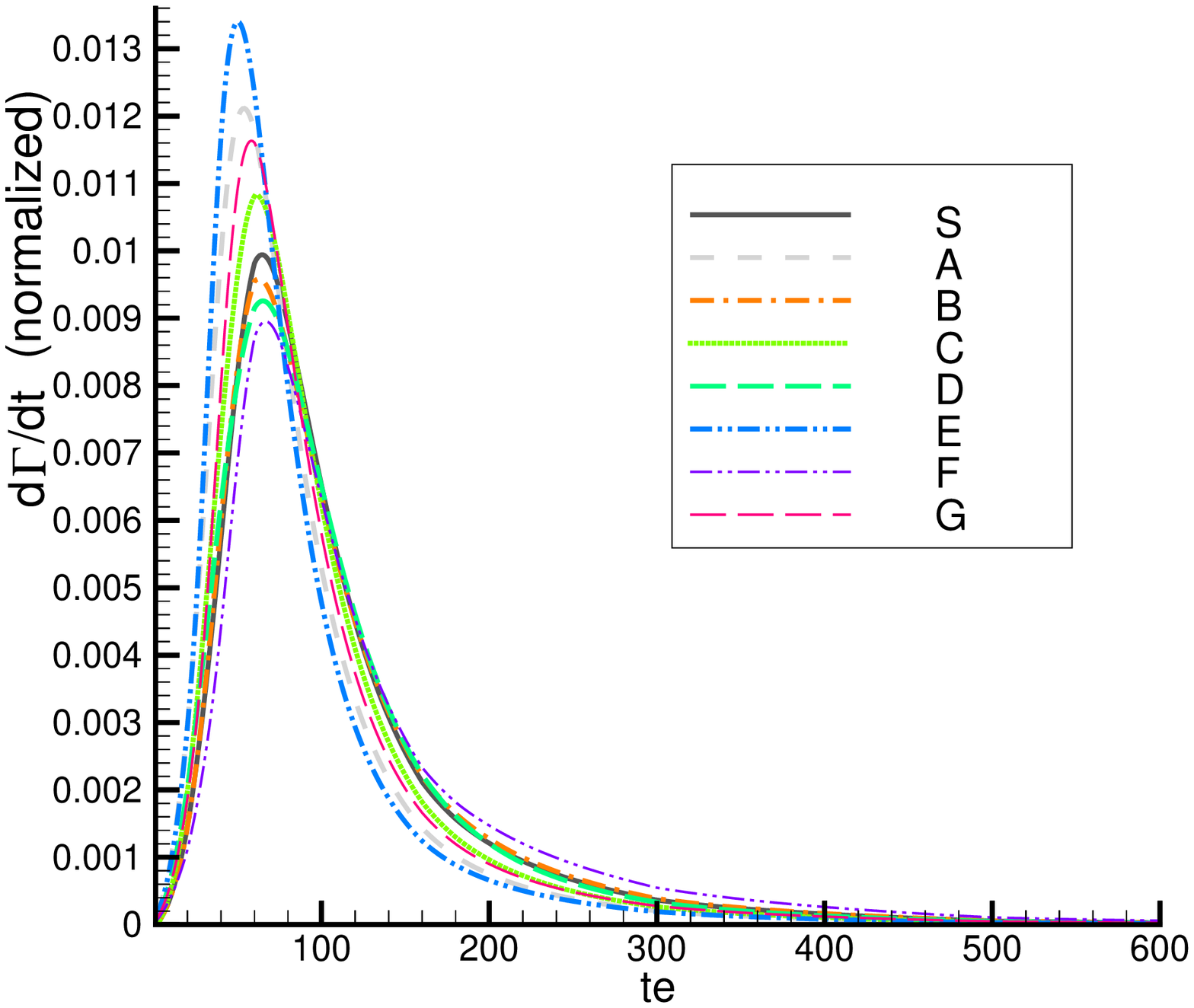,angle=0,width=8.cm,clip=}
\psfig{file=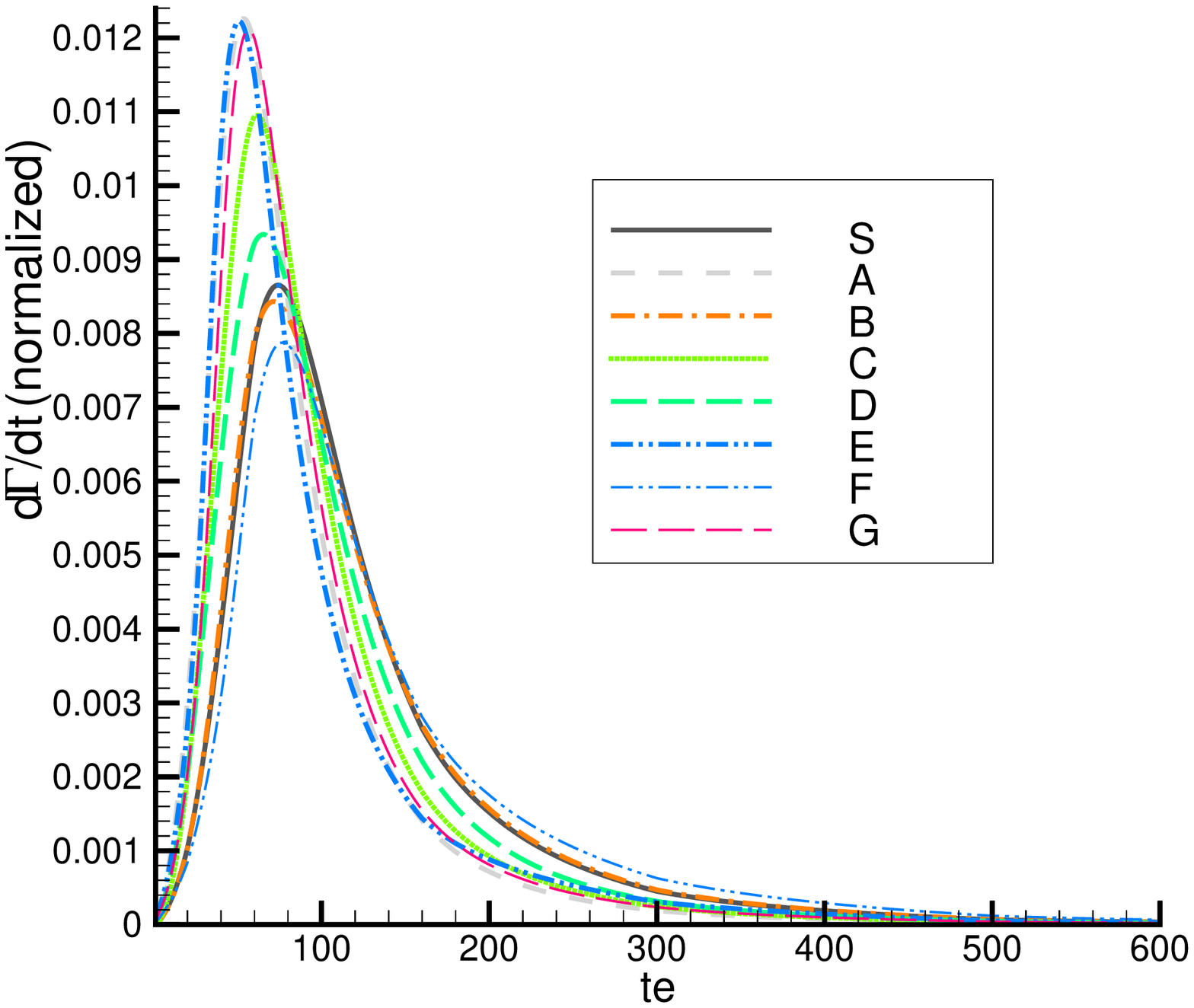,angle=0,width=8.cm,clip=} \caption{
Theoretical differential rate of events as a function of the Einstein
crossing time $t_E$ in the models described in Table 1. All
the distributions are normalized to the total number of events in
each model. According to likelihood analysis the mass function
and halo fraction depend on the microlensing candidates that have 
been obtained by the criteria A or B.
The left and right panels show the distributions of events 
that used the results of the A or B candidates, respectively
(Alcock et al. 2000, 1997).}
 \label{fig1}
\end{center}
\end{figure}
In order to deduce the expected distribution we need to have a 
reasonable knowledge of the detection efficiency of the experiments. The
detection efficiency for individual events depends on many factors
such as the impact parameter $u_0$, the moment of minimum impact
parameter $t_0$, the duration of the event $t_e$, stellar magnitude of
the lensed star, the strategy of observation and the weather conditions.
Averaging over the parameters one can obtain the efficiency as a 
function of the duration of events $\epsilon(t_e)$. The observational
efficiencies of the EROS and MACHO experiments are given in (Alcock et
al. 2000; Spiro \& Lasserre 2001). Since in MACHO experiment two
different and independent selection criteria have been used, we also
use in our study two efficiencies called A and B according to the
name of the criterion. The distribution of the rate of events expected 
from a Galactic model is obtained by multiplying the theoretical 
distribution by the observational efficiency:
\begin{equation}
\frac{d\Gamma}{dt}(expected) = \epsilon(t)\times
\frac{d\Gamma}{dt}(model).
\end{equation}
Fig.2 shows the expected distribution of the rate of events  
by applying the EROS, MACHO A and MACHO B efficiencies.
\begin{figure}
\begin{center}
\psfig{file=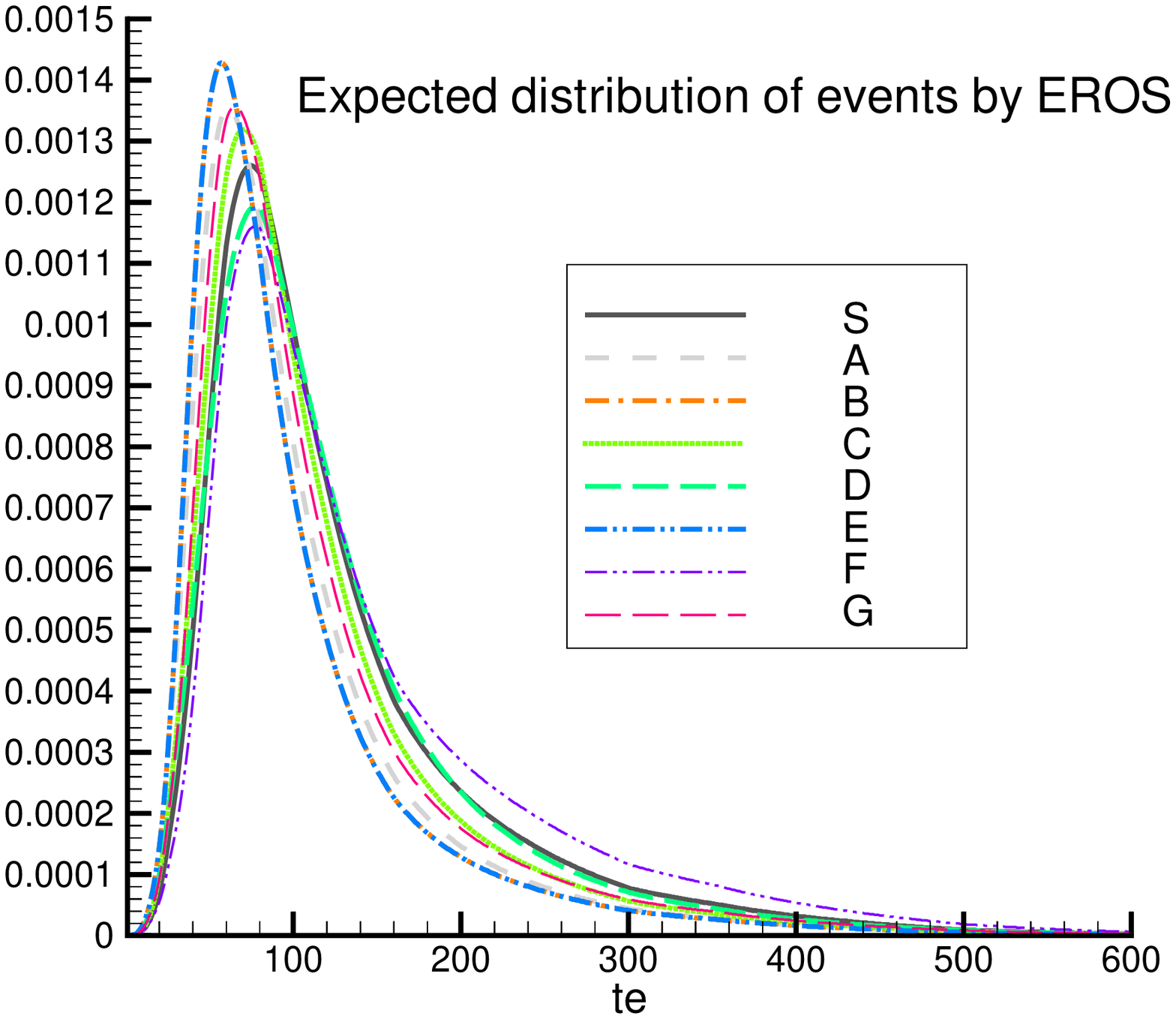,angle=0,width=8.cm,clip=}
\psfig{file=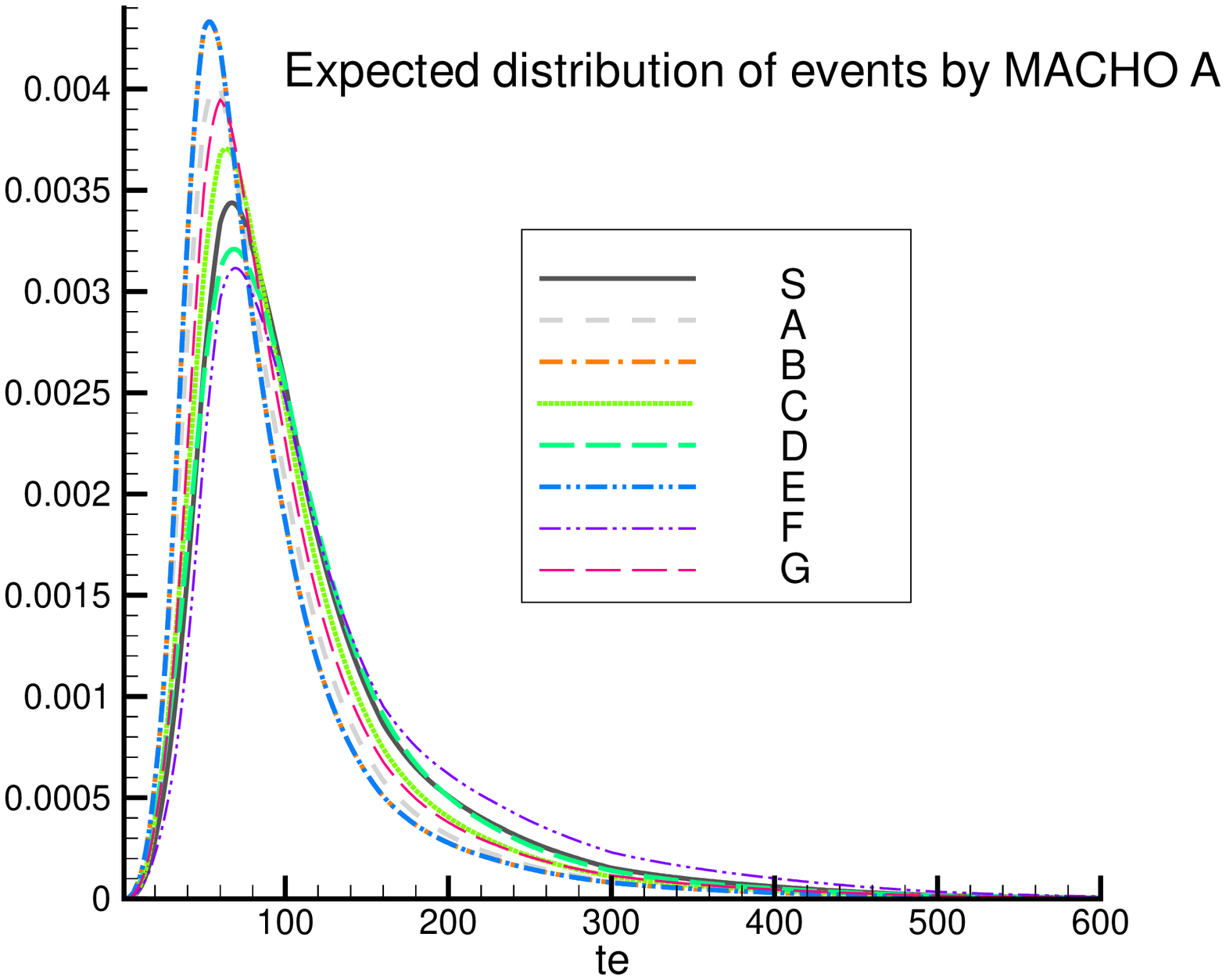,angle=0,width=8.cm,clip=}
\psfig{file=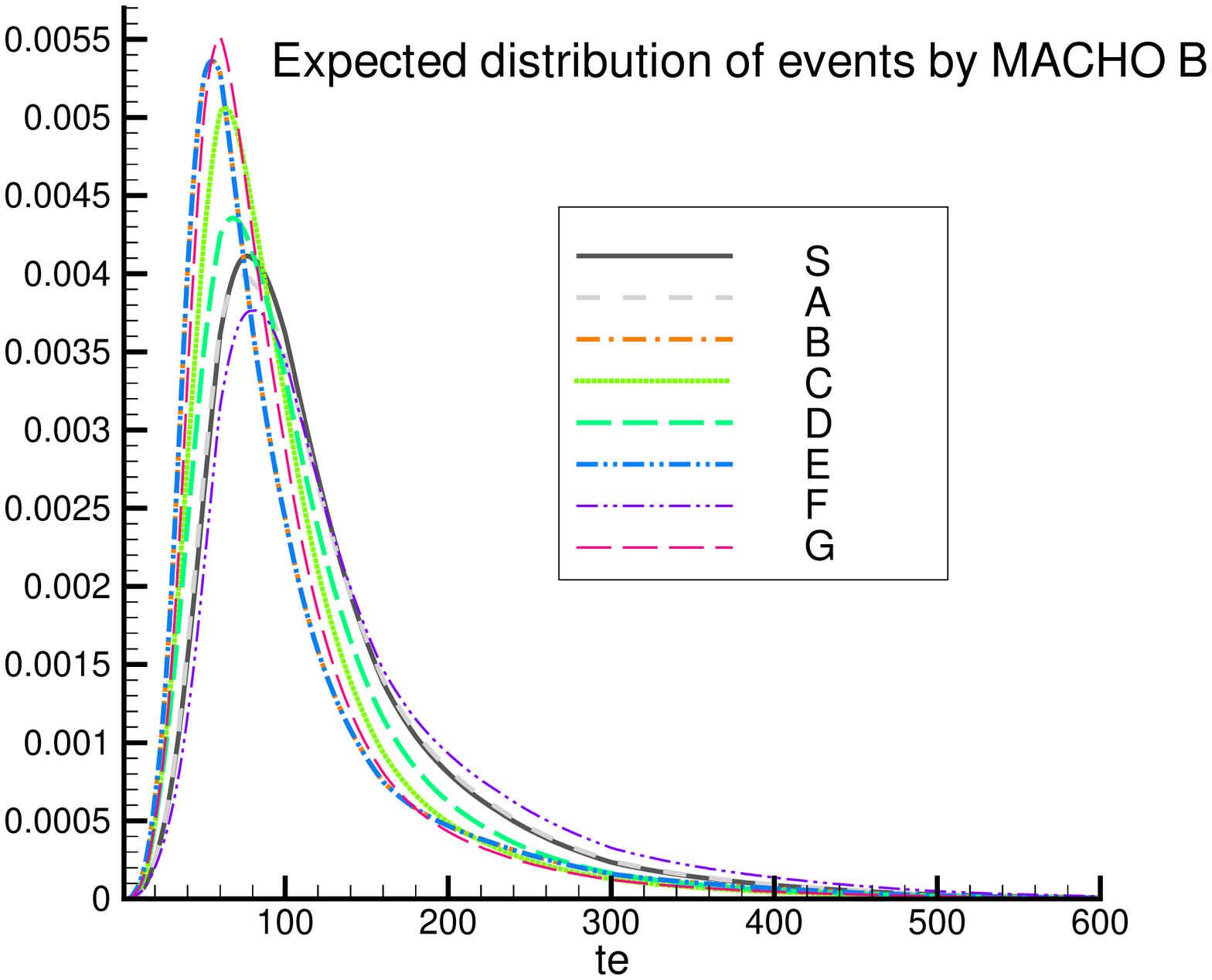,angle=0,width=8.cm,clip=}
 \caption{ The expected normalized distributions of events shown in 
various Galactic models of $S, A, B, C, D, E, F$ and 
$G$. These distributions are obtained by multiplying the theoretical 
distributions of events to the observational efficiency.}
\label{fig2}
\end{center}
\end{figure}
\section{ Microlensing Candidates and Comparison with galactic models}
In this section, the aim is to introduce the microlensing
candidates observed by the EROS and MACHO groups and find the most
likely models compatible with the data. Tables 3 and 4 show the 
microlensing candidates of the EROS and MACHO groups in terms of 
the duration of events in the direction of the LMC
\footnote{It should be mentioned that the definition of the duration
of events by the MACHO group is twice that of EROS. Here we use the 
MACHO convention.}.
The number of candidates depends on which of the criteria A or B have
been applied in the algorithm of the data reduction process. Event 22
from the MACHO candidates seems likely to be a supernova of
exceptionally long duration or an active galactic nucleus in a galaxy 
at redshift $z=0.23$, so it is ruled out as a microlensing candidate 
(Alcock et. al. 2001).
\begin{table*}
\begin{center}
\begin{tabular}{|c|c|c|c|c|c|}
\hline
Event :& LMC-1 & LMC-3 & LMC-5 & LMC-6 & LMC-7 \\
\hline
$t_E$ (days) & 46& 88 & 48 & 70 & 60 \\
\hline
\end{tabular}
\end{center}
\caption{ EROS microlensing candidates in terms of
the duration of events.}
 \label{erostab}
\end{table*}
\begin{table*}
\begin{center}
\begin{tabular}{|c|c|c|c|c|c|c|c|c|c|c|c|c|c|c|c|c|c|}
\hline
$Event :$     &1   & 4     & 5  & 6   & 7    & 8 & 9    &13  & 14  & 15 & 18 & 20& 21 &  23  &25 & 27 \\
\hline
$t_E$ (days) B& 44.& 59 & 98 & 118 & 133 & 86 & 143  &130 & 130& 47 & 96 & 94   & 121 & 110 &110 & 65 \\
$t_E$ (days) A& 41 & 55 & 92 & 112 & 125 & 81 & --   &122 & 122& 45 & 90 & --   &113  & 104&104& -- \\
\hline
\end{tabular}
\end{center}
\caption{Microlensing candidates that have been observed by the MACHO
during 5.7 years of observing 11.9 million LMC stars (Alcock et
al. 2000). Sixteen microlensing candidate results are listed from 
the reduction process by criterion B and 13 events by criterion A. Events 
that have been selected by criterion A are included in B.
} \label{machotabb}
\end{table*}
\begin{figure}
\begin{center}
\psfig{file=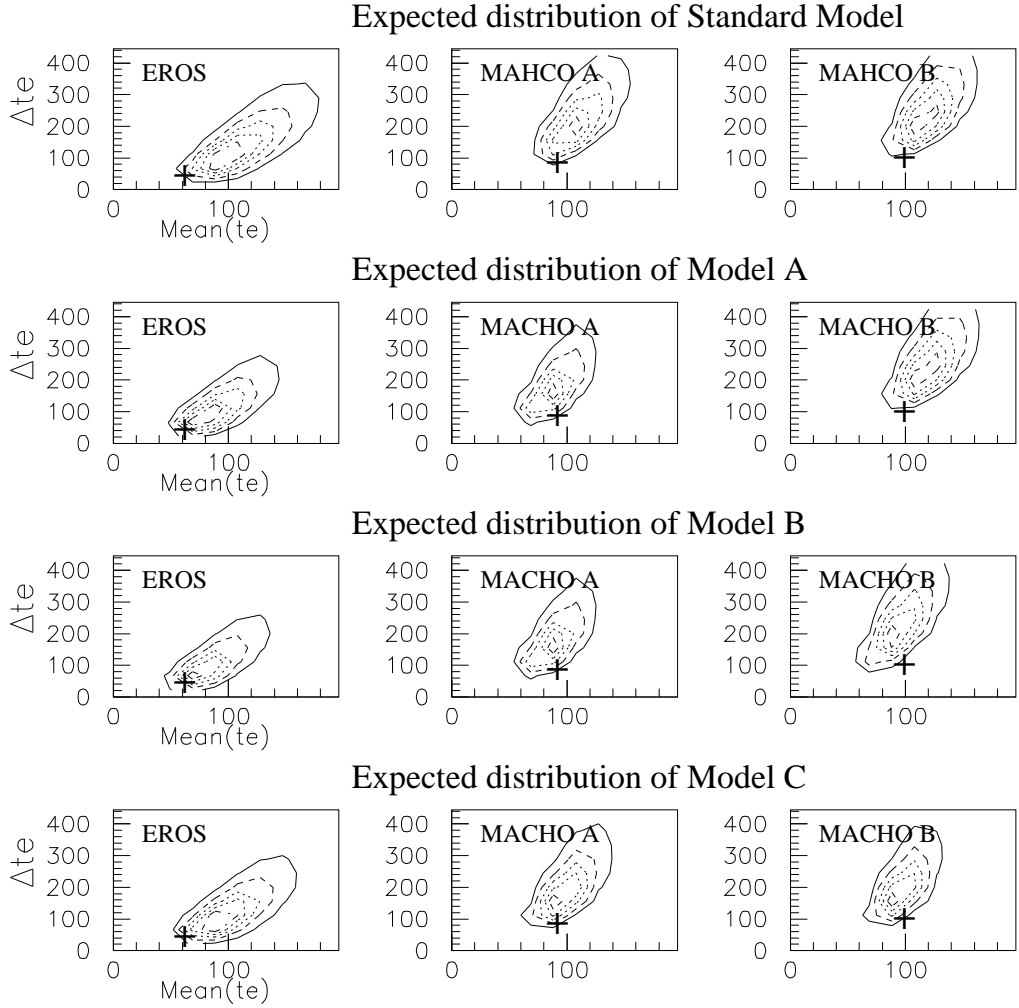,angle=0,width=15.cm,clip=} \caption{ The
expected distribution of the mean and the width of the distribution of 
the duration of events in the EROS, MACHO A and MACHO B experiments are 
shown for the standard model, and models A, B and C. The cross 
indicates the position of the observed value by these experiments. 
It is seen that non of the models gives the same width for the 
distribution of the duration or of the observed value.}
\label{fig2}
\end{center}
\end{figure}
\begin{figure}
\begin{center}
\psfig{file=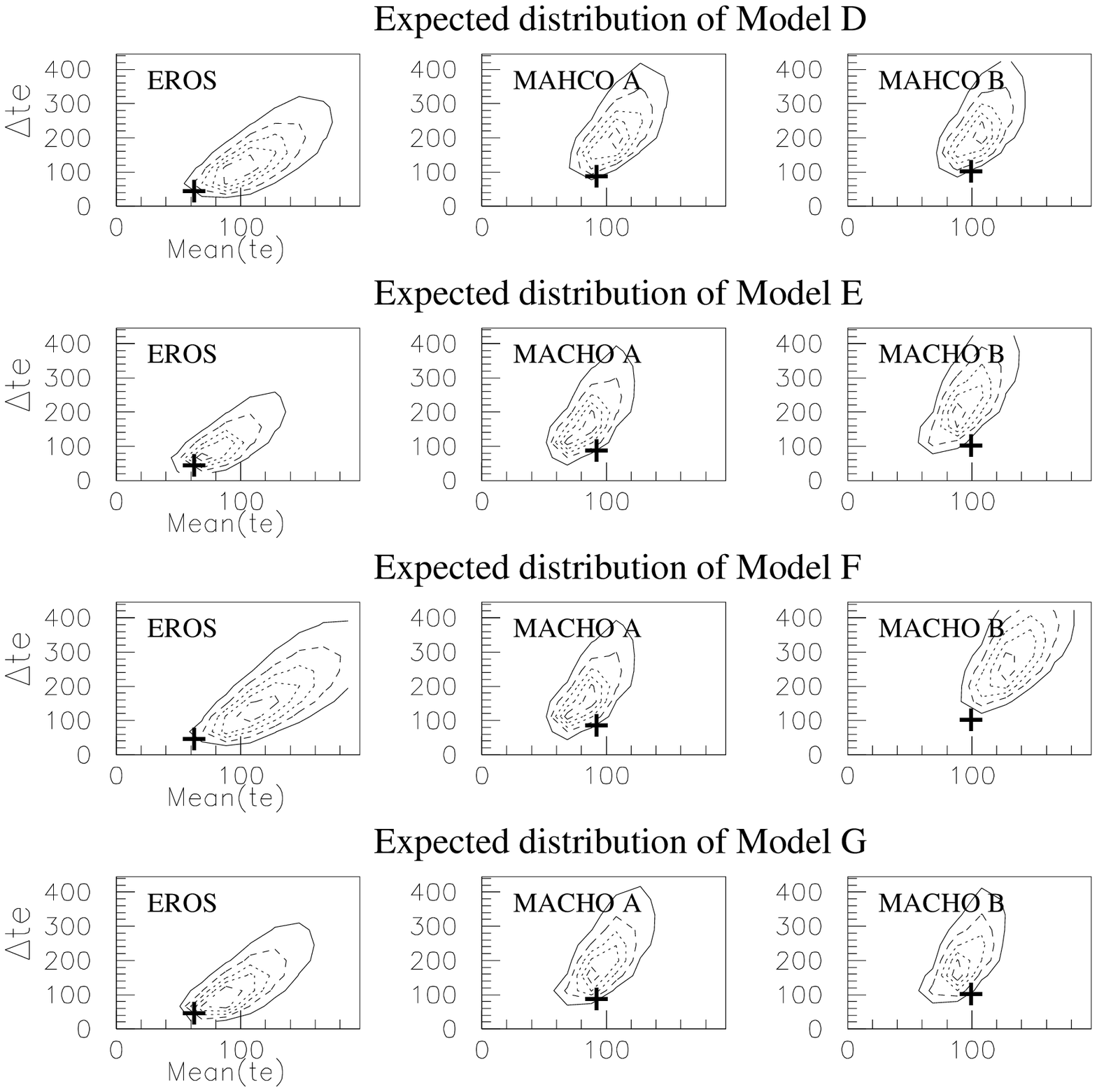,angle=0,width=15.cm,clip=}
 \caption{ The same distribution as Fig. 3 for models D, E, F and G. 
In these Galactic models also the observed value for the width of the 
duration is narrower than the expected value.}
\label{fig2}
\end{center}
\end{figure}
To compare the distribution of data in terms of the duration of events
with the expected distribution in the Galactic models, we use the two
statistical parameters called the the mean and the width of
the distribution of events (Green and Jedamzik 2002). The width of 
the duration of events for the $N_{obs}$-th observed candidate is 
defined by \footnote{The definition of the width of the duration of 
events by Green and Jedamzik has an extra normalization factor of 
the average of events duration of events 
with respect to ours. Since both definitions are 
proportional to each other, the results are unlikely to depend 
significantly on the definition.}
$$ \Delta t_E = Max_{[j = 1,N_{obs}]}(t_j) - Min_{[j=1,N_{obs}]}(t_j).$$
We note that, considering the contribution 
of non-halo lenses, these statistical parameters depend on the 
best-fitting MACHO halo fraction and the mass function 
of the halo MACHOs. In the case of EROS candidates, the mean and the width 
of events according to Table 4 are obtained as $62$ and $42$ d, 
respectively. These parameters for the MACHO candidates, from Table 5, are 
$92$ and $84$ d for criterion A, and $99$ and $99$ d for criterion 
B. We perform a Monte Calro simulation which generates the
distributions of the width and the mean of the duration of events 
from the expected theoretical distribution in the large sets of events  
where each set contains the same number of events from
the observations. In other words, the number of events in each set is 
chosen to be equal to the number of candidates from
an experiment. We chose five events in each set for generating
EROS microlensing events and $13-16$ events for the MACHO events.
Fig. 3 shows the two-dimensional distributions in terms of
the width and the mean of the duration of events in the standard model and
in models A, B and C. The crosses indicate the observed value of
the mean and the width of the distribution of duration of microlensing 
candidates in the experiments. Fig. 4 shows the same distributions for 
models D, E, F and G. Since the typical mass function and the halo 
fraction in each
model are taken from the likelihood analysis of the MACHO collaboration,
the mean of the duration of candidates is compatible with what is
expected from the models, while in all the diagrams it seems that
the width of the observed value is much narrower than the
expected distribution. To quantify this comparison, we obtain the 
fraction of generated
event samples that yield a width smaller than the observation.
Since in generating the microlensing events we take into account
background events, we compare all the observed events 
and do not reject any of them as background.
The result of this procedure, the fraction of simulations which
have smaller width compared to observation in different
Galactic models, are shown in Table 5. This fraction is less than 
about $5$ per cent in all models which means that the observed data, at 
least at the $95$ per cent level of confidence, are not compatible with 
the models.
\begin{table*}
\begin{center}
\begin{tabular}{|c|c|c|c|c|c|c|c|c|}
\hline
  & S & A  & B & C & D & E & F & G  \\
\hline
EROS Exp.     & 2  & 4  & 5 & 4  & 3 & 6 & 2   & 4       \\
MACHO A Exp.  & 1  & 3  & 4 & 1.5 & 1  & 4 & 0.5  &2.5       \\
MACHO B  Exp. & 0.5 & 0.5 & 2  & 2.5 & 1  & 2  & 0.01 &3    \\
\hline
\end{tabular}
\end{center}
\caption{The probability (per cent) that the 
the width the distribution of the duration of the observed candidates 
is smaller than that expected by 
the Galactic models. The first row indicates the name of the Galactic 
model and the first column shows the name of experiment. This 
probability is obtained by projecting the two-dimensional contour in  
Fig. 3 and 4 on the width axis and comparing the resulting distribution 
with the observed value. It is seen that the widths of the observed events 
are much smaller than the expected values from the models.}
\label{erostab}
\end{table*}
\section{Conclusion}
In this paper we have shown that in eight different Galactic
models for the Milky Way, there is discrepancy between the expected
distribution of microlensing events in terms of the duration of events
and the data from the microlensing experiments. According to the 
likelihood analysis of the MACHO collaboration, two 
parameters have been obtained from the comparison of the models with the
microlensing data: (i) the typical MACHO mass and (ii) the fraction of  
the halo mass in the form of MACHOs.
To obtain the distribution of the duration of events, we used their
results to generate the distribution of microlensing events in these 
halo models and added also the contribution of the non-halo 
components such as the disc, spheroid and LMC. After applying the 
observational efficiency the expected distributions of events were 
obtained.\\
We performed a Monte Carlo simulation to find the expected width
of the distribution of the duration and showed that the observed
width of the duration of candidates is smaller than that
expected from the standard model (Green and Jedamzik 2002). We have 
shown that the same results are also valid for the non-standard models 
of the Milky Way. The contribution of the "known" non-halo lenses in our
calculation showed that this discrepancy may not be due to background 
events. One way to explain such a narrow
distribution is that it could be due to the clumpy structure of MACHOs 
with small intrinsic velocity dispersion along the line of sight. If this
were the case, the expected distribution of duration should be narrow 
compared to the ordinary halo case. The other advantage of this model
could be decreasing that the mean mass of the MACHOs decrease compared to 
$\sim 0.5 M_{\odot}$ .\\
The blending effect also changes our estimation of the actual value for 
the duration of events. The next generation of microlensing experiments, 
such as SUPERMACHO (Stubbs 1999) surveys with better sampling of
microlensing light curves and high photometric precision, on the one
hand, and increasing in the number of candidates, on the other, should 
reduce the ambiguity due to Poisson statistics. One of the
proposed projects is to use two telescope working together,
the first one to detect the microlensing events, and a follow-up 2-meter 
class telescope to observe the events precisely
(Rahvar {\it el al} 2003). This type of survey could also partially break 
the degeneracy between the lens parameters by parallax,
finite-size and double lens effects to localize the position and
identify the mass of the lenses.

\end{document}